# Topological Hall effect arising from the mesoscopic and microscopic non-coplanar magnetic structure in MnBi

Yangkun He[1,*], Sebastian Schneider[2,3], Toni Helm[1,4], Jacob Gayles[1,5], Daniel Wolf[3], Ivan Soldatov[3,7], Horst Borrmann[1], Walter Schnelle[1], Rudolf Schaefer[3,6], Gerhard H. Fecher[1], Bernd Rellinghaus[2], Claudia Felser[1]

[1]*Max-Planck-Institute for Chemical Physics of Solids, D-01187 Dresden, Germany*
[2]*Dresden Center for Nanoanalysis (DCN), Center for Advancing Electronics Dresden (cfaed), TU Dresden, 01062 Dresden*
[3]*Leibniz Institute for Solid State and Materials Research (IFW) Dresden, Helmholtz strasse 20, D-01069 Dresden, Germany*
[4]*Dresden High Magnetic Field Laboratory (HLD-EMFL), Helmholtz-Zentrum Dresden–Rossendorf, 01328 Dresden, Germany*
[5]*Department of Physics, University of South Florida, Tampa, Florida 33620, USA*
[6]*Institute for Materials Science, TU Dresden, D-01062 Dresden, Germany*
[7] *Institute of Natural Sciences and Mathematic, Ural Federal University, Yekaterinburg 620075, Russia*

**The topological Hall effect (THE), induced by the Berry curvature, which originates from non-zero scalar spin chirality, is an important feature for mesoscopic topological structures, such as skyrmions. However, the THE might also arise from other microscopic non-coplanar spin structures in the lattice. Thus, the origin of the THE inevitably needs to be determined to fully understand skyrmions and find new host materials. Here, we examine the Hall effect in both bulk- and micron-sized lamellar samples of MnBi. The sample size affects the temperature and field range in which the THE is detectable. Although bulk sample exhibits the THE only upon exposure to weak fields in the easy-cone state, in thin lamella the THE exists across a wide temperature range and occurs at fields near saturation. Our results show that both the non-coplanar spin structure in the lattice and topologically non-trivial skyrmion bubbles are responsible for the THE, and that the dominant mechanism depends on the sample size. Hence, the magnetic phase diagram for MnBi is strongly size-dependent. Our study provides an example in which the THE is simultaneously induced by two mechanisms, and builds a bridge between mesoscopic and microscopic magnetic structures.**

The topological Hall effect (THE) in systems with non-zero scalar spin chirality has recently attracted considerable attention[1]. In non-coplanar spin textures, the THE is caused by an emerging magnetic field accompanied by Berry curvature[2,3], which is created when an electron acquires a Berry phase either (i) by travelling through the non-zero topological winding of the spin texture of a skyrmion or a skyrmion bubble (**Figure 1**a), or (ii) when it hops along a non-collinear spin configuration that provides for a finite scalar spin chirality (Figure 1b). In the case of skyrmions, the length scale of the spin texture is sufficiently longer than the underlying lattice constant in the first case ($10^0$ to $10^3$ nm), whereas it is equal to the lattice constant in the second case ($10^{-1}$ to $10^0$ nm).

Because the THE is a hallmark of magnetic skyrmions, its detection is broadly used as a characterization tool to identify new host materials that contain skyrmions[4], which are highly interesting in that they could find application as memory and logic elements in future computing devices[5,6]. Probing the THE by magnetotransport measurements thereby offers the advantage of gaining insights into a system without the need for expensive studies by neutron diffraction[7] or low-temperature Lorentz transmission electron microscopy (L-TEM)[8]. As a result, a large number of materials with the THE were reported and announced to be candidates for skyrmion materials based only on the presence of the THE as detected by magnetotransport measurements[9,10]. However, only a few of these materials were finally confirmed to host skyrmions or topological spin structures[11], and in most of these materials, this was the result of non-coplanar spin structures in the lattice. On the other hand, it remains unclear why in certain skyrmionic materials, although they were identified as hosting skyrmions by L-TEM, the THE is observable, but emerges at different temperatures[12,13] or in different field directions[14,15] depending on the sample thickness. These discrepancies show that the detection of the THE alone does not provide unambiguous proof for the presence of the effect with either of the two origins mentioned above. Ways in which to distinguish the origin of the THE would therefore need to be found.

Skyrmions are stabilized by the Dzyaloshinskii–Moriya interaction (DMI) inherent to magnetic materials that lack inversion symmetry. Alternatively, disruption of the inversion symmetry can be engineered at interfaces. Other topologically non-trivial magnetic mesoscopic structures, such as Type-I bubbles, were observed in centrosymmetric systems[16]. In non-centrosymmetric skyrmion materials these bubbles are mainly stabilized by dipole-dipole interactions rather than a strong DMI. Another important difference between bubbles and skyrmions is that the former have a uniformly magnetized core whereas the latter do not. However, their topology is similar in the following respects: In Type I bubbles, the magnetization in the Bloch-type domain wall surrounding the core of the bubble circulates either clockwise or anticlockwise. With a winding number of $N_{\text{win}} = 1$[17], these bubbles are topologically equivalent to Bloch skyrmions with the skyrmion number $N_{\text{sk}} = 1$[5] therefore, they are known as "skyrmion bubbles"[18]. Because they have the same topological properties, a THE can also arise from these bubbles.

One prominent example of a centrosymmetric magnet is MnBi. It crystallizes in the centrosymmetric NiAs-type structure of the space group $P6_3/mmc$ (No. 194) with alternating Mn and Bi layers. The lattice constants are $a$ = 4.2876(5) Å and $c$ = 6.1154(5) Å based on our X-ray diffraction results. Known as a room-temperature hard magnet with a large magnetic moment of 3.94 $\mu_B$/Mn and a high Curie temperature ($T_c$ > 630 K)[19], MnBi has a magnetocrystalline anisotropy $K_1$ of 1 MJm$^{-3}$ at room temperature[20], sufficient to resist self-demagnetization into the in-plane domains of thin lamella and to stabilize bubbles. Below the spin-orientation transition temperature $T_{SR1}$ = 137 K, the magnetic moment gradually rotates away from the *c*-axis upon cooling[21] until it abruptly locks into the *ab*-plane[19] at $T_{SR2}$ = 87 K. In this study, we selected MnBi to investigate the THE for two reasons: i) Similar to other highly anisotropic magnets (e.g., L1$_0$ Fe-Pt[22] and barium ferrite[23]), the strong uniaxial anisotropy might possibly lead to a real-space topological structure and the existence of the THE; ii) The easy-cone structure between $T_{SR2}$ and $T_{SR1}$ is non-coplanar and could also give rise to the THE.

In this work, we examined the THE in MnBi both in bulk and thin, micron-sized samples. In the micron-sized samples, we observed topologically non-trivial (Type I) skyrmion bubbles in MnBi above 87 K using L-TEM and off-axis electron holography. The corresponding THE appears close to the saturation point. In the bulk samples, the maximum THE appears at low fields, below 137 K, as a result of the non-coplanar spin structure in the easy-cone state. Our study provides an excellent example of approaches that could be followed to distinguish the THE induced by either mesoscopic or microscopic non-coplanar magnetic structures and may help to identify new skyrmion materials.

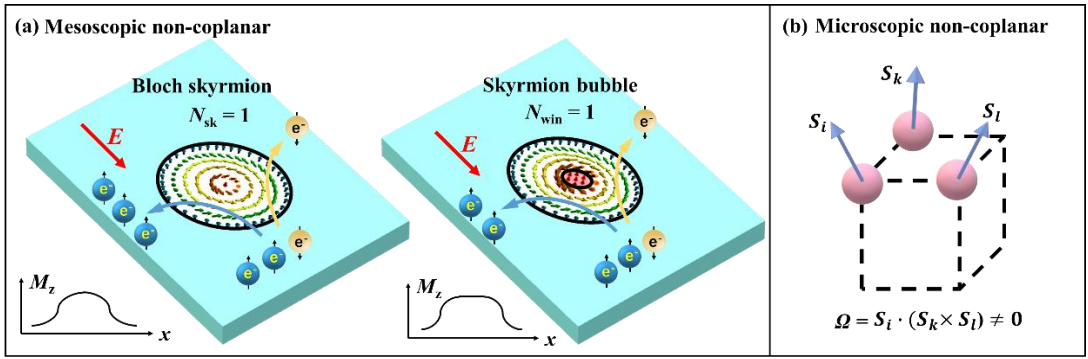

**Figure 1. Two mechanisms responsible for THE.** (a) THE induced by the non-zero topological winding of the spin texture in real space. Spin polarized electrons are scattered in different directions in Bloch skyrmions and skyrmion bubbles, causing the THE. The inset shows a schematic profile of the out-of-plane magnetization $M_z$ in the cross section of the structures along the in-plane direction *x*. (b) THE induced by microscopic non-coplanar spin structure. Schematic view of the non-coplanar spin arrangement $S_i$, $S_k$, and $S_l$ in a lattice, with the solid angle, $\Omega$.

**Figure 2**a and 2b shows the Hall resistivity of bulk MnBi measured at 130 K and 200 K, respectively. The Hall resistivity observed in magnetic material may be attributed to the ordinary Hall effect $\rho_{OHE}$ due to the Lorenz force, the anomalous Hall effect $\rho_{AHE}$, or the topological Hall effect $\rho_{THE}$. The total Hall resistivity is simply a sum of these contributions:

$$\rho_{xy} = \rho_{OHE} + \rho_{AHE} + \rho_{THE} \qquad (1)$$

The magnitudes of the resistivity $\rho_{OHE}$ and $\rho_{AHE}$ are proportional to the magnetic field and the magnetization, respectively. At 130 K in the fitted curve for $\rho_{THE} = \rho_{xy} - \rho_{OHE} - \rho_{AHE}$ with a positive magnetic field, a dip corresponding to the maximum (black line) of 0.013 μΩ cm, appears at a low field of 0.18 T during both the magnetization and demagnetization processes. Evidence of the THE at 200 K was not found.

In the micron-sized device, we observed the THE over a large range of temperatures from 130 K to 200 K, as shown in Figure 2c. The Hall signal remains constant at low fields, after which it abruptly intensifies, followed by a continuous increase towards saturation. Rather than being linear, this increase is characterized by additional irregular behavior (THE) from approximately 0.4 to 0.6 T depending on the temperature. However, these irregularities were not observed to exist in the magnetization data of the same crystal in supplementary information of which the shape is plotted as $\rho_{AHE}$. In Figure 2d, we fitted the data at 200 K using Eq. 1. A maximum topological Hall resistivity of 0.021 µΩ cm appears near saturation upon exposure to an increasingly strong magnetic field. Note that the THE does not manifest itself during demagnetization in the same field, which means that the THE is sensitive to the magnetic history. At 300 K, the micron-sized device undergoes a transition into a hard magnet indicated by a square hysteresis loop (supplementary information). In this phase, the THE is not observable.

The magnetization curve and Hall resistivity at 130 K are shown in Figure 2e and 2f, respectively. At low fields, the coercivity in the Hall measurements is much larger than that for magnetization, indicating that $\rho_{xy}$ is dominated by $\rho_{THE}$ rather than $\rho_{AHE}$. Note that the upward (green) sweep is above the downward (orange) sweep at low field values, whereas this is not the case for the magnetization. This behavior only exists below $T_{SR1}$ = 137 K, where the magnetization has an easy-cone structure. The topological Hall resistivity remains non-zero until saturation with large hysteresis.

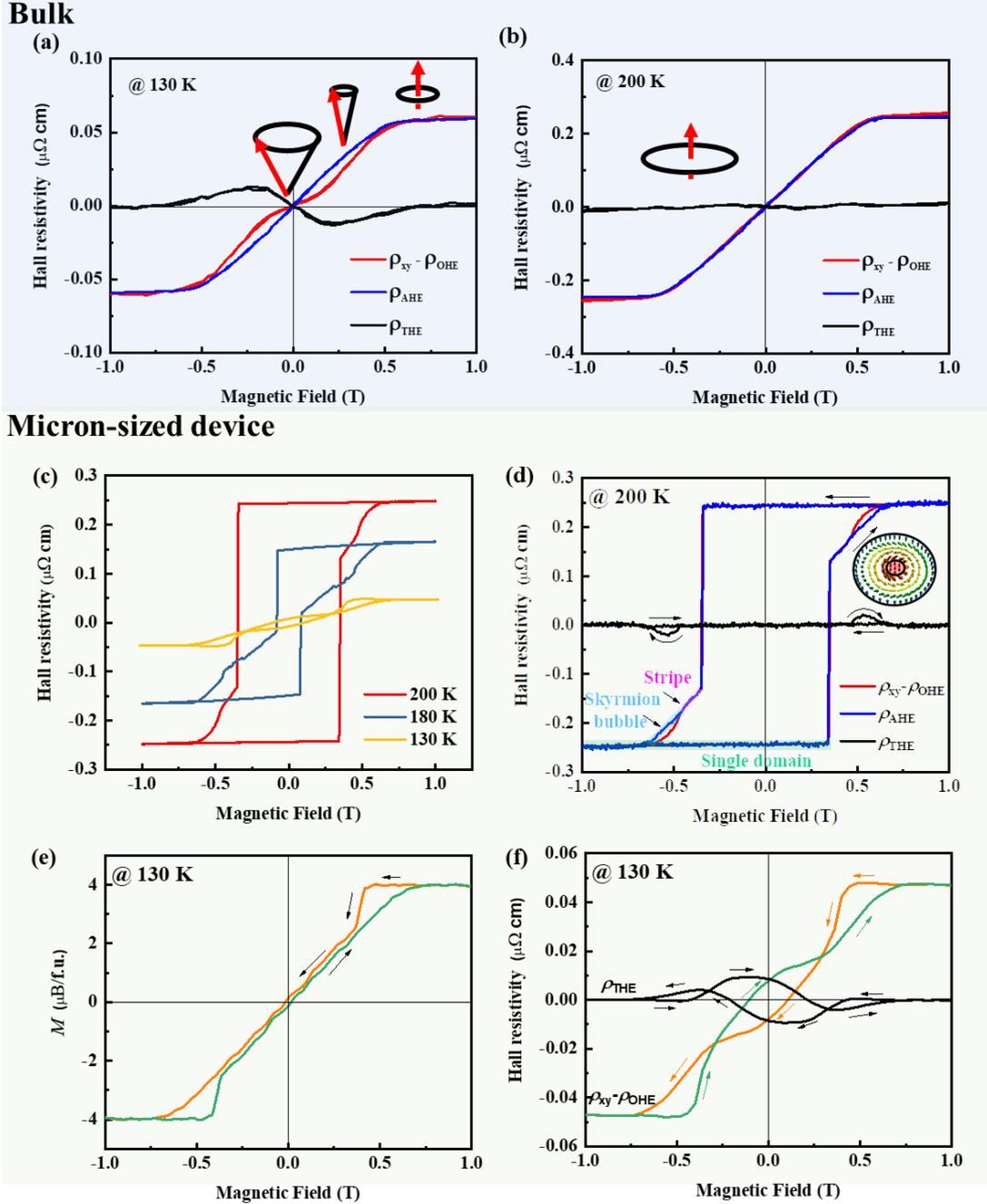

**Figure 2. THE in bulk and micron-sized devices.** The current runs along the *a*-direction and the external field is oriented parallel to the *c*-direction of the crystal. THE for a bulk sample (approximately 500 μm thick) at (a) 130 K and (b) 200 K. The inset shows the spin structure at the corresponding temperature. The cone angle shrinks during magnetization at 130 K until the system is transformed into the collinear state along the *c*-axis at saturation. (c) Hall effect for a micron-sized sample (cross-section: 1.16 μm thick, 2.24 μm wide) at different temperatures. The additional irregularities at approximately 0.4 to 0.6 T are the result of an additional THE contribution. (d) Curves fitted using Eq. 1 for the micron-sized device at 200 K. The fields in which the domain structure changes from a single domain to stripe domains (sometimes called band domains[24]), to skyrmion bubbles, and finally to a single domain at saturation are shown in different colors (see L-TEM experiments). (e) Magnetization curves at 130 K. The magnetization and demagnetization processes are indicated in green and orange, respectively. (f) The $\rho_{THE}$ (black line) and $\rho_{xy}$-$\rho_{OHE}$ at 130 K. Note that the sign of the difference between the upward (green) and downward (orange) sweep changes at low-field values, whereas that of the magnetization does not. The same sample was used to measure both the magnetization and Hall resistivity of the device.

To gain further insight into the origin of the difference in the THE observed for bulk and micron-sized MnBi, we studied the magnetic properties and domain structure in more detail. In **Figure 3**, we compare the field-dependent magnetization of bulk MnBi for the magnetic field aligned in-plane and out-of-plane at room temperature and 2 K. The magnetization at 300 K M saturates faster for $B \parallel c$ than for $B \parallel a$, and the opposite occurs at 2K, confirming the spin-reorientation transition observed by neutron studies[19,21]. Two transition events are discernible in our AC-susceptibility measurements, the first occurring at $T_{SR1}$ = 137 K and the second at $T_{SR2}$ = 87 K. At $T > T_{SR1}$, MnBi is an easy-axis magnet; at $T_{SR2} <$ T $< T_{SR1}$, an easy-cone phase is established, and when T $< T_{SR2}$, MnBi transforms into an easy-plane phase. The transitions into the different magnetic phases are also observable in the susceptibility deduced from the magnetization curves, as shown in Figure 3c. In the easy-axis region, the moment is parallel to the $c$-axis with a two-phase branched domain[24] structure observable at the (001) surface (see Figure 3d), typical for a highly anisotropic uniaxial magnet. In the magnetization process, these branched domains persist until saturation (see supplementary information). Therefore, non-coplanar spin configurations are absent for $T > T_{SR1}$, and hence the THE is not detected. However, for $T_{SR2} <$ T $< T_{SR1}$, the conical spin structure is non-coplanar, giving rise to the THE. The cone angle is the largest at zero field and shrinks gradually as the magnetization intensifies, as shown in Figure 2a; therefore, the maximum THE occurs at low field.

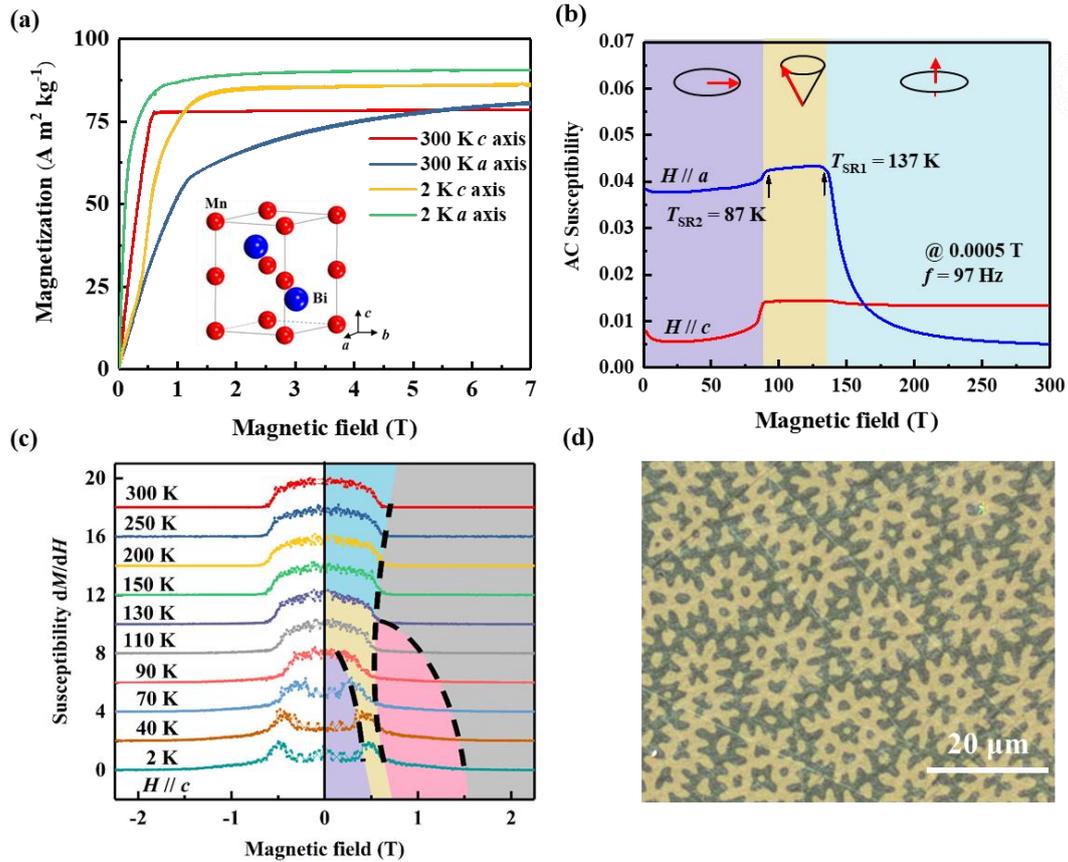

**Figure 3. Magnetic properties and domain structure in bulk MnBi.** (a) Magnetization curve along the *c*- and *a*-axes at 2 K and 300 K. (b) AC susceptibility measurements. (c) Susceptibility along the *c*-axis at different temperatures and fields obtained from the magnetization curves. For improved visualization, offsets of 2 were added. (d) Branched domains at the (001) surface obtained by Kerr microscopy at zero field and room temperature.

As we reduced the thickness, the dipole–dipole interaction increasingly influenced the domain structure (supplementary information). This structure in the thin lamella of MnBi was observed by L-TEM at different temperatures with $B \parallel c$, as shown in **Figure 4**. At room temperature, MnBi is a hard magnet. Therefore, it maintains a single-domain structure once magnetized. Upon cooling, the anisotropy decreases and the stripe domains appear because of the dipole–dipole interaction below 200 K. An increase in the magnetic field causes the stripe domains parallel to the field to expand, whereas those that are antiparallel to the field shrink. Finally, the stripe domains disintegrate into individual bubbles.

Additionally, we employed off-axis electron holography to reveal the direction of the projected in-plane magnetic induction of the bubbles shown in Figure 4f. We observed bubbles that exhibit a Type I structure, with the spin continuously rotating clockwise within the plane. The spin chirality of this structure is 1, which is topologically equivalent to Bloch-type skyrmions and is expected to induce the THE. Other types of magnetic bubbles, such as topologically trivial Type II bubbles, were not observed[16]. The magnetic structure appears to be robust even against an additional in-plane magnetic field, achieved by tilting the lamella by 5 degrees[18,23]. The domain wall is approximately 50 nm wide, and the central area of which the moment is oriented perpendicular to the plane has a diameter of 100 nm. Therefore, approximately 75% of the volume of the bubble consists of a non-collinear spin structure (the remaining 25% is attributed to the core), which is not far from 100%. This is the typical signature of skyrmions. The magnetic fields at which the skyrmion bubbles are observed correspond to the field at which the THE is detected in our transport measurement in Figs. 2c and 2d. Apparently, the THE in the micron-sized devices is induced by domain wall scattering in skyrmion bubbles. After the saturation point in Figure 4d, during the demagnetization process, skyrmion bubbles do not emerge again. Instead, the sudden appearance of stripe domains (see Figure 4e), which remain in the low-field region, corresponds with the magnetization and Hall resistivity data. This provides an explanation for the hysteresis loop shown in Figure 2e. Note that in both cases, that is, below and above $T_{SR1}$, magnetic bubbles do exist, regardless of whether the magnetization of the bulk materials is characterized by an easy axis or an easy cone. Below $T_{SR2}$, MnBi prefers an in-plane domain structure, and the stripe domain and skyrmion bubbles disappear.

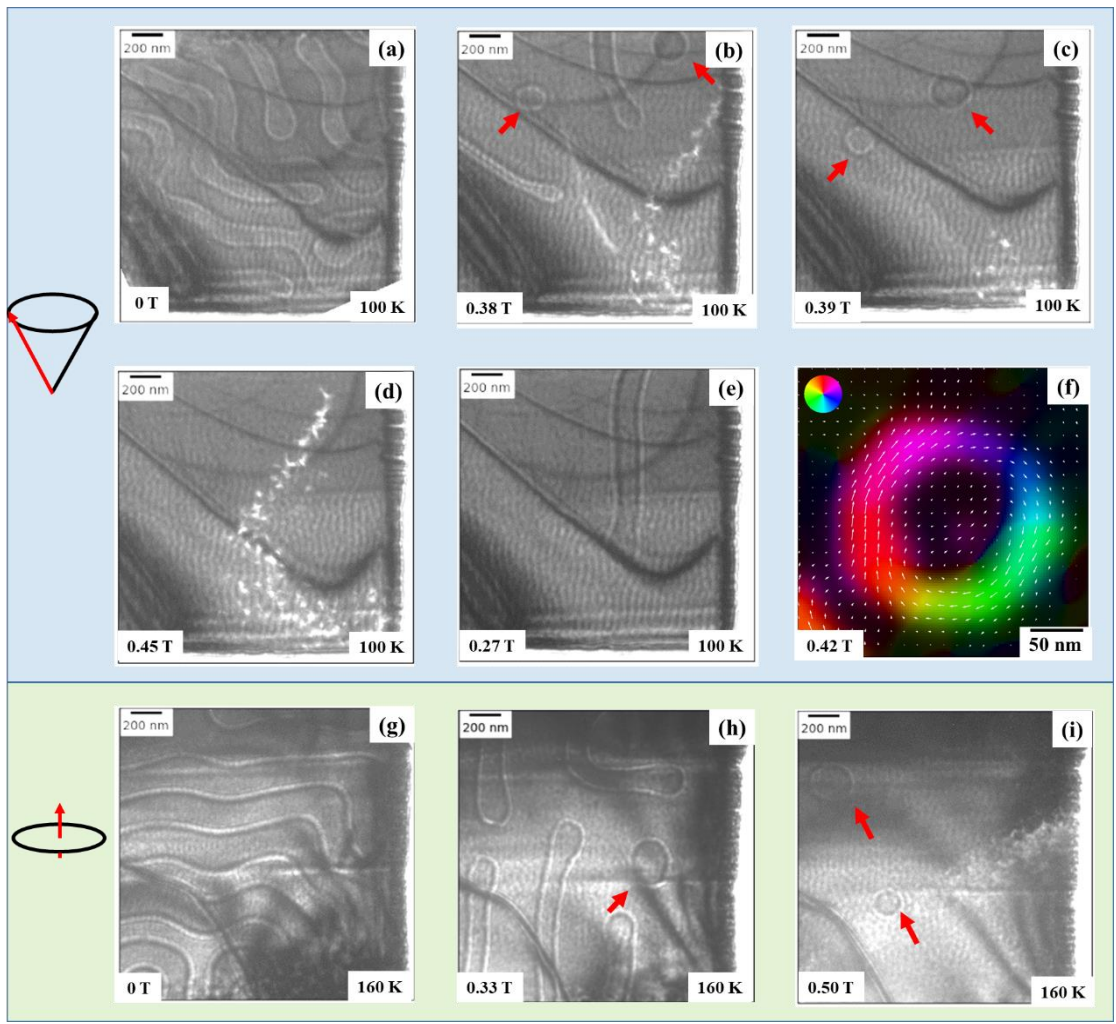

**Figure 4. Characterization of the spin texture with L-TEM and off-axis electron holography.**

(a)-(e) Domain structure observed at 100 K. (a) Stripe domains. (b) Stripes + bubbles. (c) Type I bubbles. (d) Single domain. (e) Stripe domain after demagnetization from saturation without bubbles. (f) Direction of the projected in-plane magnetic induction. (g)-(i) Domain structure observed at 160 K. Skyrmion bubbles with a diameter of approximately 100 nm are indicated by red arrows.

**Figure 5**a presents a schematic phase diagram of *bulk* MnBi. This is based on the susceptibility measurements for $B \parallel c$ at different temperatures, as shown in Figure 3c. Upon exposure to a magnetic field, the easy-cone structure extends to lower temperatures. Skyrmion bubbles were not observed across the entire range of the magnetic field and temperature. The domain phase diagram for *the micron-sized* samples is shown in Figure 5b and is based on the THE and images of the skyrmion bubbles. The green area corresponds to the field and temperature range in which we observed skyrmion bubbles in MnBi. At higher temperatures, the system behaves as a hard magnet with a single domain structure owing to strong magnetocrystalline anisotropy. At lower temperatures, the in-plane multi-domain state is the ground state.

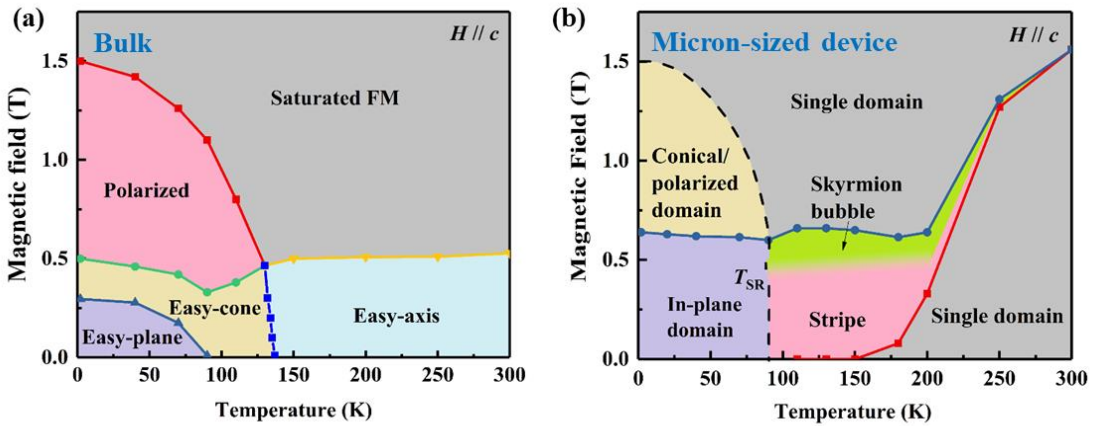

**Figure 5. Schematic magnetic phase diagram of (a) bulk and (b) micron-sized device**. The data points were obtained from our studies of the magnetic susceptibility, magnetotransport and from L-TEM microscopy. Here the polarized state refers to a transitional state from saturation to an easy-cone state.

The THE, observed in our micron-sized samples at 130 K, can now be explained by the coexistence of two mechanisms. Although a similar phenomenon of zero-field THE at low field was reported before and skyrmions were expected[25,26], we did not detect any evidence of mesoscopic topological spin structures at zero field after demagnetization in our L-TEM images, as shown in Figure 4e. Note that the phenomenon only exists below 137 K in the easy-cone phase. Because the spin is only slightly canted away from the *c*-axis, the in-plane stray field, the driving force for domain wall motion, is small. Therefore, the hysteresis in the topological Hall resistivity could be related to the hysteresis in the magnetization along the *ab* plane. The topological Hall resistivity for a micron-sized device at 130 K can be understood to change from the bulk THE with additional effects including the hysteresis and the THE due to skyrmion bubbles near saturation. The third term is responsible for reducing the size of the dip at approximately 0.4 T in the magnetization process compared with that at approximately 0.2 T in the demagnetization process, as shown in the supplementary information. Note that we can exclude the possibility of the skyrmion Hall effect[27] because our current density is 38 A cm$^{-2}$, which is 4–5 orders too low to move the skyrmion bubbles.

In our work, we show that it is possible to distinguish whether the measured THE originates from either mesoscopic skyrmions (bubbles in our case) or microscopic non-coplanar spin structures. The domain structure of materials that contain skyrmions and bubbles at zero field consists of helicoidal or stripe domains, and the skyrmions are stabilized by an applied magnetic field. Thus, the THE originating from the real-space topology should not appear close to zero field in MnBi. On the other hand, the THE induced by non-coplanar spin structures in the lattice is expected to emerge at low field. These two conclusions are not limited to MnBi but can also be extended to many other systems. Indeed, the existence of the THE close to saturation was similarly observed in $DyCo_3$ films[28] and Co-Ir-Pt multi-layered films[29], where skyrmions were observed. In bulk materials such as $Fe_5Sn_3$[30], MnP,[31] and $Cr_5Te_8$[32], the THE was found at low fields owing to the non-coplanar spin structure.

In conclusion, we determined the THE we observed in bulk and mesoscopic samples of MnBi to have different origins. The THE in bulk samples is closely connected to the non-coplanar easy-cone spin structure between 87 and 137 K. The cone angle shrinks with increasing magnetic field; therefore, the maximum THE appears in the low-field region. For thin micron-sized devices, the THE is related to the presence of non-trivial skyrmion bubbles across a large temperature range above 87 K. The magnetic domain structure transforms from stripe domains to bubbles near the saturation field; therefore, the THE appears at high field. Specifically, the THE that exists at 130 K in the mesoscopic samples, could be attributed to two coexisting mechanisms. Our conclusion can be extended to other high-anisotropy uniaxial materials in which skyrmions/bubbles exist, and could be useful for identifying the mechanisms responsible for the THE.

**Methods**
**Single-crystal characterization** Single crystals of MnBi were grown by the flux method[19]. Wavelength dispersive X-ray spectroscopy was used to determine the composition of the crystals to be homogenous and existing of $Mn_{49.2}Bi_{50.8}$ with an error of approximately 0.5 %. The crystals were characterized by powder X-ray diffraction as being single-phase with a hexagonal structure. The orientation of the single crystals was confirmed by the Laue method.
**Preparation of micron-sized transport device** Thin lamellae of MnBi were cut using a focused ion beam (FIB). The sample size was 17.10 (*a*-axis) × 2.24 (*b*-axis) × 1.16 (*c*-axis) µm$^3$. The lamellae were affixed to sapphire substrates by using epoxy and gold contacts that were deposited via sputtering.
**Magnetization measurements.** The magnetic properties were measured on single crystals using a vibrating sample magnetometer (MPMS 3, Quantum Design). The bulk sample size was approximately 1.5 mm in diameter (*ab*-plane) and 0.5 mm in thickness (*c*-axis). In the case of measurements of the thin lamella, an empty sample holder containing only the substrate and epoxy was measured for comparison to remove the diamagnetic background.

**Magneto-optical Kerr microscopy** Domain images were obtained using the polar Kerr effect in a wide-field magneto-optical Kerr microscope[33] at room temperature with a polished single crystal with a thickness of 0.5 mm and a (001) surface.

**Electrical transport measurements** The Hall resistivity was measured in a Quantum Design PPMS 9 T instrument by applying a standard four-probe method. The current for the bulk and micron-sized samples were 16 mA and 1 mA, respectively.

**Lorentz transmission electron microscopy and off-axis electron holography.** The magnetic phases were imaged with a double-corrected FEI Titan³ 80–300 microscope operated in the corrected Lorentz mode at an acceleration voltage of 300 kV. The sample was maintained at the required temperature by using liquid nitrogen in a Gatan double-tilt cooling holder. The off-axis electron holography experiment was conducted at a biprism voltage of 200 V.

## Acknowledgements

This work was financially supported by an Advanced Grant from the European Research Council (No. 742068) "TOPMAT," the European Union's Horizon 2020 research and innovation programme (No. 824123) "SKYTOP," the European Union's Horizon 2020 research and innovation programme (No. 766566) "ASPIN," the Deutsche Forschungsgemeinschaft (Project-ID 258499086) "SFB 1143," the Deutsche Forschungsgemeinschaft (Project-ID FE 633/30-1) "SPP Skyrmions," the DFG through the Würzburg-Dresden Cluster of Excellence on Complexity and Topology in Quantum Matter ct.qmat (EXC 2147, Project-ID 39085490). I.S. is grateful to Deutsche Forschungsgemeinschaft for supporting this work through project SO 1623/2-1.

## Author contributions

Single crystals were grown by Y.H. The crystal was characterized and the magnetic and transport measurements were performed by Y.H. with the help of H.B. and W.S. L-TEM and off-axis electron holography experiments were performed by S.S. and D.W. B.R. co-supervised the TEM investigations. The FIB microstructure transport devices were fabricated by T.H. Kerr microscopy images were taken by I.S. and R.S. J.G. provided theoretical support. All authors discussed the results. The paper was written by Y.H. and G.H.F. with feedback from all the authors. The project was supervised by C. F.

## Competing financial interests

The authors declare no competing financial interests.

## Keywords

Topological Hall effect, skyrmion bubble, MnBi, noncoplanar spin structure